%% file: srsd.tex
\documentclass[letterpaper,11pt]{report}

\pdfoutput=1

\usepackage{graphicx}
\usepackage{latexsym}
\usepackage{makeidx}
\usepackage{url}
\usepackage{hyperref}

\makeindex

\input{commands}

\begin{document}
\pagestyle{headings}

\title
{
	Ftklipse -- Design and Implementation of an Extendable Computer Forensics Environment\\
	Software Requirements Specification Document
}

\author{Marc-Andr\'e Laverdi\`ere \and Serguei A. Mokhov \and Suhasini Tsapa \and Djamel Benredjem}

\date{April 2006}
\maketitle

\include{introduction}
\include{description}
\include{requirements}
\include{references}
\include{support-info}



\end{document}

%% file: commands.tex
\newcommand{\xf}[1]{Figure~\ref{#1}}

\newcommand{\lucidL}[1]{{$\mathit{Lucid}$}($L$) }

		{}

\def\myvert{\raise 2.27pt \hbox{\vrule depth 0pt height 8pt width 0.2mm}}
\def\myarrow{\hspace*{0.43mm}%
             \raise 2.29pt\hbox{\vrule depth 0pt height 8pt width 0.16mm}%
             \hspace*{-0.32mm}%
             $\longrightarrow$
             \ %
             }


%% file: introduction.tex
\chapter{Introduction}
\index{Introduction}

\section{Purpose}

The purpose behind this document is to describe the features of
ftklipse, an extendable platform for computer forensics.
This document will explain the product for the customer, as well
as provide a detailed specification for the developer.

\section{Scope}

Ftklipse is a thick-client solution for forensics investigation.
It allows to collect and preserve evidence, to analyze it and to report
on it.

It supports chain of custody management, access control policies and batch
operation of its included tools in order to facilitate and accelerate the
investigation. The environment itself and its tools are configurable as well.

\section{Definitions and Acronyms}

\begin{description}
  \item[Cryptographic Hash Function] Function mapping input data of an arbitrary size to
     a fixed-sized output that is highly collision resistant.
  \item[JVM] The Java Virtual Machine. Program and framework allowing the execution of program
     developed using the Java programming language.
  \item[GUI] Graphical User Interface.

\end{description}

\section{Compliance}
This document was written based on \cite{standards98ieee}.



%% file: description.tex
\chapter{Overall Description}

\section{Product Perspective}

\begin{itemize}
  \item Ftklipse is meant to be a stand-alone product, depending on a variety of standard tools organized as plug-ins.
  \item Ftklipse is meant to be extendable using plug-ins that will add evidence gathering and analysis properties
  \item The product has only one interface, a graphical user interface residing on the client computer
\end{itemize}

\subsection{System interfaces}
The only interface to the system will be its GUI.

\subsection{User Interfaces}
Ftlipse implements a user interfaces that is evidence-centric. It offers wizards for each of its
features for ease of use. It allows investigators to record notes for each piece of evidence as well
as to record additional reporting information. Please refer to \xf{fig:show:intro} and \xf{fig:show:evidence} for
an example of the look and feel of the application.

\begin{figure}
\includegraphics[scale=0.4, angle=90, keepaspectratio=true]{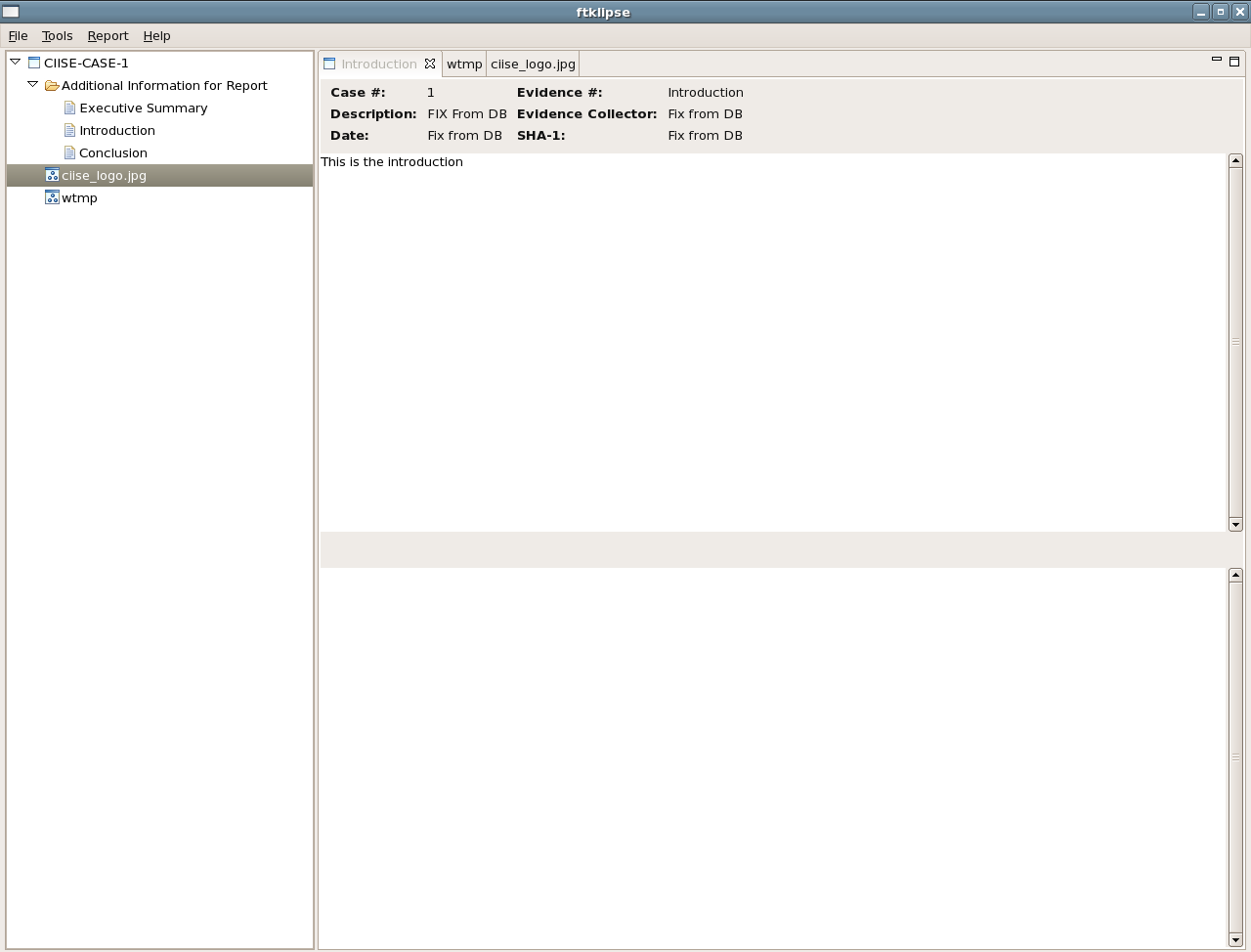}
\caption{User Interface Showing the Case Introduction}
\label{fig:show:intro}
\end{figure}

\begin{figure}
\includegraphics[scale=0.4, angle=90, keepaspectratio=true]{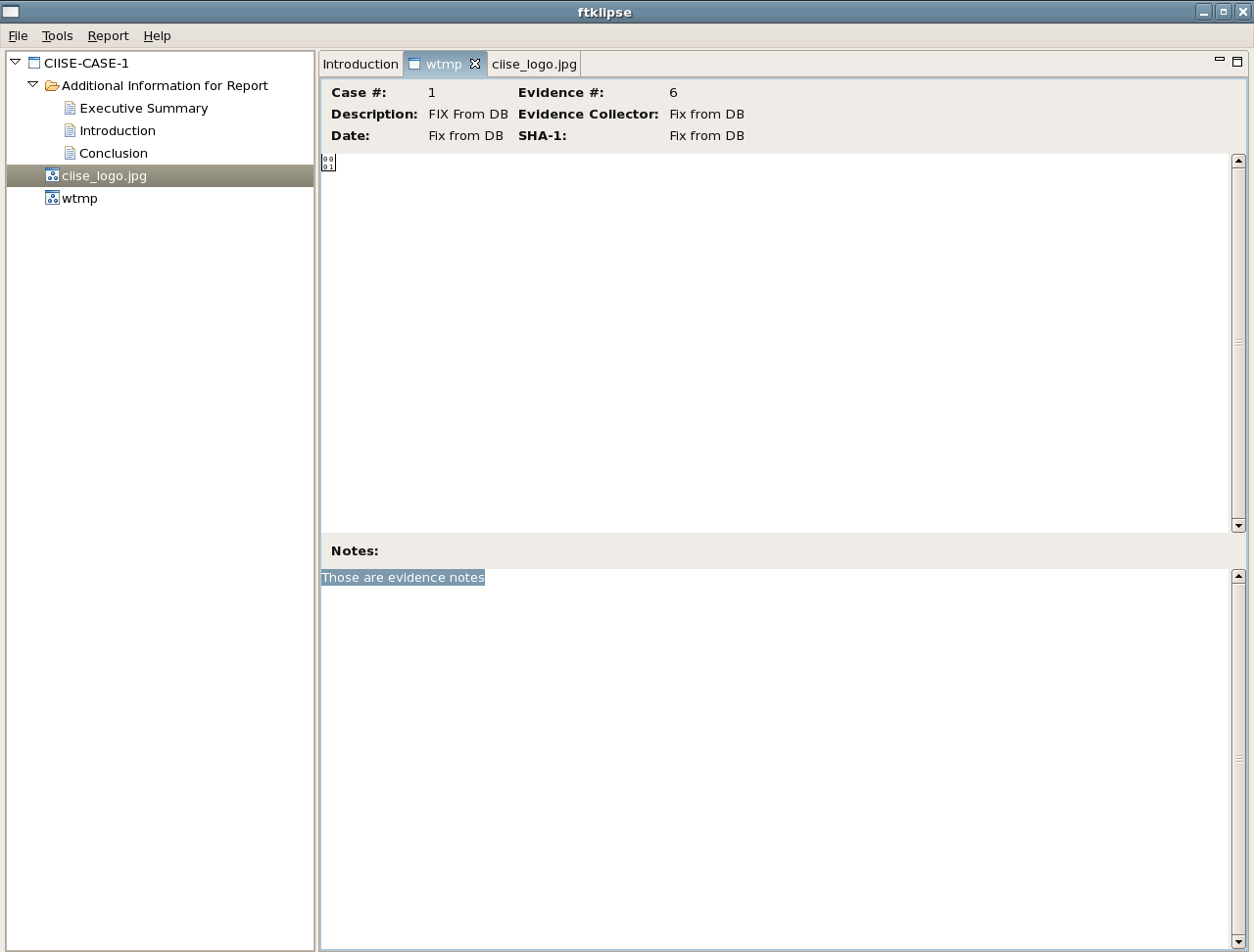}
\caption{User Interface Showing the Evidence Information and Notes}
\label{fig:show:evidence}
\end{figure}

\subsection{Software Interfaces}
The product must expose a software interface for plug-in developers to use.
The interfaces provided must allow to:
\begin{itemize}
  \item Register the plug-in
  \item Extend the Graphical User Interface's tool menus (window, pop-up, etc.)
  \item Offer an interface for the plug-in to implement to allow callbacks enabling execution
\end{itemize}

\section{Product Functions}

The system will implement the following functionalities:
\begin{itemize}
  \item Creation of cases
  \item Evidence Gathering using integrated and plug-in tools
  \item Evidence Integrity validation using a hash function
  \item Evidence Import from any media to an existing case
  \item Logging of all operations performed on the evidence
  \item Validation of integrity of evidence after each operation over it
  \item Display of evidence in read-only mode either in ASCII, Unicode or Hex formats
  \item Recording of investigative notes for each piece of evidence
  \item Capability to extract a part of the evidence into another file
  \item Capability to copy and rename the copy of the evidence
  \item Generation of reports in PDF and \LaTeX2e formats that includes listing of the evidence in the case,
	a printout of selected parts of the evidence, the investigative notes related to selected parts of
	the evidence and a customized executive summary, introduction, and conclusion. It also integrates
	the chain of custody information for each part of the evidence displaying the principal, time stamp
	and operation performed on the evidence.
  \item An extendable set of tools through a plug-in architecture
  \item Tool-specific defaults and configuration screens
\end{itemize}
\section{User Characteristics}

Users are cyber forensics investigators. They are experienced using existing sets of tools,
and will be trained in the use of ftklipse before its deployment.

Indirect users are investigators, prosecutors, judges and laypersons, which will consult the
reports generated. They expect reports of high quality which demonstrate objectivity and methodology.

\section{Constraints}

\subsection{Hardware Constraints}
Any computer able to operate the Eclipse platform can be used to operate Ftklipse.

\subsection{Software Constraints}
It is assumed that the investigator's computer supports and includes the following programs:
\begin{itemize}
  \item JVM, version 5 or higher
  \item \LaTeX2e, preferably pdflatex
\end{itemize}

Other tools are not assumed to be present, as they are integrated in each plug-in.

In the case of using Ftklipse for evidence collection only, only the JVM is required.

\section{Assumptions and Dependencies}

The software assumes a non-hostile environment (i.e. not aiming at disturbing its operation).

\section{Apportioning of requirements}

Some features are to be implemented in later versions of Ftklipse, notably:
\begin{itemize}
  \item Integration of the Access Control framework with administrator screens
  \item \LaTeX output of reports
  \item Object-specific logging
  \item Hexadecimal and image display
  \item Evidence Extraction
\end{itemize}


%% file: requirements.tex
\chapter{Specific Requirements}

\section{External Interfaces}
The product must expose a software interface for plug-in developers to use.
The interfaces provided must allow to:
\begin{itemize}
  \item Register the plug-in
  \item Extend the Graphical User Interface's tool menus (window, pop-up, etc.)
  \item Offer an interface for the plug-in to implement to allow callbacks enabling execution
\end{itemize}

\section{Functional Requirements}

\subsection{Domain Model}

Our domain model is a traditional police investigation one, augmented with some
information specific to cyber forensics and our requirements\cite{debbabi}. It is summarized in \xf{fig:dom}.
\begin{figure}
  \centering
  \includegraphics{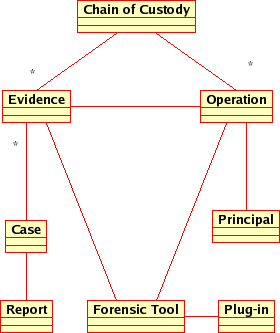}
  \caption{Domain Model for Ftklipse}
  \label{fig:dom}
\end{figure}

\subsection{Use Case Model}
The use case model for Ftklipse is illustrated in \xf{fig:use_case_diagram}.
\begin{figure}
  \includegraphics[width=\textwidth, keepaspectratio=true]{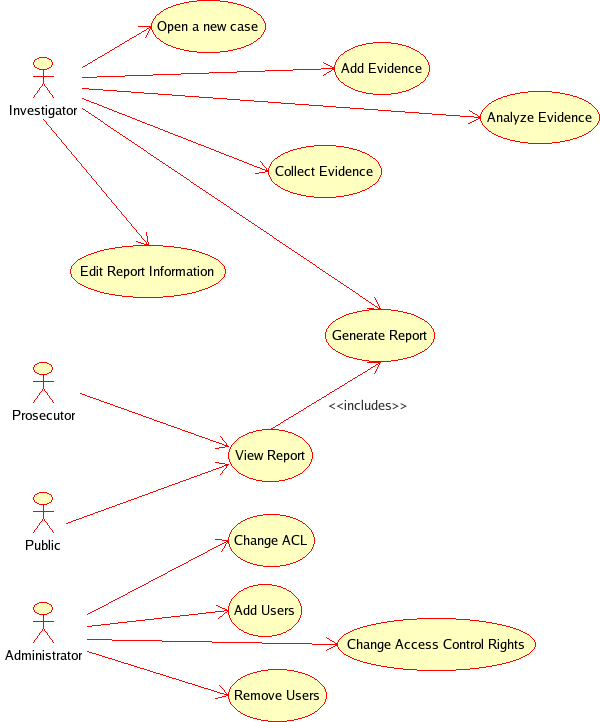}
  \caption{Use Case Diagram for Ftklipse}
  \label{fig:use_case_diagram}
\end{figure}

\section{Requirements Description}

  \subsection{Creation of cases}
 	\paragraph{Description}
Ftklipse allows the creation of cases with their associated metadata, as specified in section \ref{sec:db_ref}.
	\paragraph{Criticality}
This feature is critical to the software
	\paragraph{Technical Issues}
None
	\paragraph{Dependencies with Other Requirements}
None

  \subsection{Evidence Gathering}
 	\paragraph{Description}
Ftklipse allows to run different tools in order to perform evidence collection on a live system.
	\paragraph{Criticality}
This feature is critical to the software.

	\paragraph{Technical Issues}
The collection of the output of the gathering tool can be problematic, considering the variety
of tools and their working. The redirection of the tool's standard input and output in a manner
useful to the investigator should be considered.
	\paragraph{Dependencies with Other Requirements}
None

  \subsection{Evidence Analysis}
 	\paragraph{Description}
Ftklipse allows to run different tools on one or more selected evidences, as well as to 
operate a batch analysis. In the latter case, the system must offer a GUI to the user
that allows the selection of the evidence and operations to perform on it.

	\paragraph{Criticality}
The ability to analyze the evidence is critical. However, the automated analysis of multiple
pieces of evidence is not critical.
	\paragraph{Technical Issues}
The development of a generic programming interface for the variety of analysis tools
is likely to be complex.
	\paragraph{Dependencies with Other Requirements}
None

  \subsection{Evidence Integrity Validation}
 	\paragraph{Description}
Ftklipse records the SHA-1 signature of every piece of evidence and ensures that the evidence is
kept correct during the investigation. In the case of a corruption of the evidence, Ftklipse
detects it and records which operation caused this corruption.

	\paragraph{Criticality}
This feature is important to the operation of the software, although not critical.
	\paragraph{Technical Issues}
	\paragraph{Dependencies with Other Requirements}

  \subsection{Evidence Import}
 	\paragraph{Description}
Ftklipse allows to import evidence that was collected outside of itself. The evidence must be
accompanied by a SHA-1 digest that is correct in order to import the evidence in the system.
	\paragraph{Criticality}
This feature is important, although not critical.
	\paragraph{Technical Issues}
The encoding and format of the SHA-1 signature can vary from one tool to another.

	\paragraph{Dependencies with Other Requirements}

  \subsection{Logging}
 	\paragraph{Description}
All operations are logged globally by Ftklipse. Furthermore, all operations
related to a given piece of evidence are logged for that evidence specifically.
	\paragraph{Criticality}
The global logging is critical to Ftklipse. The specific logging is important, but not essential.
	\paragraph{Technical Issues}
	\paragraph{Dependencies with Other Requirements}

  \subsection{Evidence Display}
 	\paragraph{Description}
The evidence can be visualized, if authorized, in read-only mode either in ASCII, Unicode or Hex formats. Furthermore,
images can be viewed within Ftklipse and can be opened in an external viewer program.
	\paragraph{Criticality}
This function is critical to the operation of the software in ASCII.
	\paragraph{Technical Issues}
	\paragraph{Dependencies with Other Requirements}

  \subsection{Recording of Investigative Notes}
 	\paragraph{Description}
The investigator must be able to record information regarding each piece of evidence, as well as report-specific information.

	\paragraph{Criticality}
This function is critical to the operation of Ftklipse.
	\paragraph{Technical Issues}
	\paragraph{Dependencies with Other Requirements}

  \subsection{Evidence Extraction}
 	\paragraph{Description}
The investigator must be able to select a subset of the viewed evidence and extract it into another file,
which will then be treated as evidence itself. Ftklipse must record this operation and keep relationship
information in the database of evidence.
	\paragraph{Criticality}
This feature is of moderate importance.
	\paragraph{Technical Issues}
	\paragraph{Dependencies with Other Requirements}

  \subsection{Evidence Cloning}
 	\paragraph{Description}
The investigator must be able to copy a piece of evidence in full and optionally to rename the copy.
	\paragraph{Criticality}
This feature is nice to have.
	\paragraph{Technical Issues}
	\paragraph{Dependencies with Other Requirements}

  \subsection{Report Generation}
 	\paragraph{Description}
The investigator must be able to generate a report for a selected case that includes all evidence,
their notes, as well as other report-specific data. The output formats can be PDF or \LaTeX2e.
	\paragraph{Criticality}
This feature is critical.
	\paragraph{Technical Issues}
	\paragraph{Dependencies with Other Requirements}

  \subsection{Plug-in Architecture}
 	\paragraph{Description}
Ftklipse allows third-party developers to create plug-ins that can be added at configuration
time by system administrators.
	\paragraph{Criticality}
This feature is critical.
	\paragraph{Technical Issues}
	\paragraph{Dependencies with Other Requirements}

  \subsection{Access Control Management}
 	\paragraph{Description}
Ftklipse operates with an access control list for each case, piece of evidence, and report information.
Each user must be authenticated and each operation must be authorized in the view of the user's access
rights.

Notably, the rights that must be implemented are:
\begin{itemize}
  \item View rights over a case or piece of evidence. This defines if the user is authorized to be aware of the 
	existence of a given case or piece of evidence.
  \item Read rights over a case or piece of evidence. This defines if the user, being previously granted view
	rights over the object, is able to read the case's information or visualize or operate on a piece of evidence.
  \item Write rights over a case or piece of evidence. This defines if the user is authorized to add to the general
	case notes or the evidence notes. This also defines if the user is allowed to add evidence to a given case.
\end{itemize}

By default, Ftklipse must offer default access rights based on the user's role, as well as default access rights
for different categories of objects.

Ftklipse must provide GUI tools to manage the both user and object rights.

	\paragraph{Criticality}
This feature is important, not critical.
	\paragraph{Technical Issues}
The implementation of the access control algorithm can be complex. Furthermore, some administration functions
(such as the impact of a redefinition of default rights) require some thought to ensure that no previously
confidential information becomes publicly available.
	\paragraph{Dependencies with Other Requirements}

  \subsection{Tool-specific defaults and configuration screens}
 	\paragraph{Description}
Each tool is responsible to maintain its state, notably regarding its default settings
which must be modifiable by the user and preserved from one run of ftklipse to another.

Each tool must supply a screen that allows to set the proper parameters before the operation
of the tool.

Default options are to be used on direct invocation of the tool.

	\paragraph{Criticality}

This feature is important
	\paragraph{Technical Issues}
	\paragraph{Dependencies with Other Requirements}

\section{Performance Requirements}

Ftklipse does not have any particular performance requirements

\section{Logical Database Requirements}
\label{sec:db_ref}
A database is required in order to store the case management and chain of custody information.

The database must be able to store:
\begin{itemize}
  \item The relationship between parts of the evidence
  \item The operations done on the evidence, including its time stamp, its description and the investigator that performed it.
\end{itemize}

The information that must be tracked by the database is the following:
\begin{itemize}
  \item The case's meta-information (ID, details, description, timestamps, investigators)
  \item The case's evidence.
  \item The user credentials.
  \item The object access control lists.
  \item The chain of custody over every piece of evidence. This includes the cryptographic hash 
	sums, the operations performed on the evidence and the principal who performed it.
\end{itemize}

\section{Design Constraints}

The design must take in consideration that the base implementation language is Java.
It also must take in consideration the different options of the tools that can be plugged into it.

\section{Software System Attributes}

In this section, we describe the non-functional attributes
of Ftklipse.

\subsection{Security}

\subsection{Reliability}
The software must behave correctly during 20 continuous hours of operation.

\subsection{Availability}
There are no availability constraints.

\subsection{Maintainability}
The software must allow for tool plug-ins to be integrated automatically.
The software must also be self-updatable.

\subsection{Portability}
The software must operate on POSIX and Windows systems. Tools integrated in the software
must be adjusted accordingly.




%% file: references.tex
\addcontentsline{toc}{chapter}{References}


\bibliography{srsd}
\bibliographystyle{alpha}


%% file: support-info.tex
\chapter{Supporting Information}

\tableofcontents
\listoffigures